\documentclass[preprint,aps,showpacs,showkeys,12pt]{revtex4}
\newtheorem{theorem}{Theorem}[section]
\newtheorem{corollary}[theorem]{Corollary}
\newtheorem{proposition}[theorem]{Proposition}

\newtheorem{example}[theorem]{Example}
\newtheorem{remark}[theorem]{Remark}

\newcommand{\proof}{\noindent {\bf Proof. }}

\newcounter{acount}

\newfont{\BB}{msbm10 scaled\magstep1}
\newfont{\bb}{msbm8}
\def\R{\mbox{\BB R}}

\def\D{\mbox{\BB D}}

\def\tr{{\rm trace}\ }
\begin{document}
\title{On entropy production for controlled Markovian evolution}

\author{Michele Pavon}
 \homepage{http://www.math.unipd.it/~pavon}
\affiliation{Dipartimento di Matematica Pura ed Applicata,
Universit\`a di Padova, via Belzoni 7, 35131 Padova, Italy}
\email{pavon@math.unipd.it}
\author{Francesco Ticozzi}
\affiliation{Dipartimento di Ingegneria dell'Informazione,
Universit\`a di Padova, via Gradenigo 6/B, 35131 Padova, Italy}
\email{ticozzi@dei.unipd.it}

\date{\today}

\begin{abstract}
We consider thermodynamic systems with finitely many degrees of
freedom and subject to an external control action.  We derive some
basic results on the dependence of the relative entropy production
rate on the  controlling force. Applications to macromolecular
cooling and to controlling the convergence to equilibrium rate are
sketched.  Analogous results are derived  for closed and open
$n$-level quantum  systems.
\end{abstract}

 \maketitle

\section{Introduction}
Advances in nanotechnology permit nowadays to implement feedback
control actions on nanodevices. For instance, in surface
topography, the deflection of a cantilever is captured by a
photodetector that records the angle of reflection from a laser
beam focused on the mirrored surface on
 back side of the cantilever. Position feedback control is used to maintain the probe at a constant force or distance from the object surface. Position can also be differentiated allowing to apply a velocity dependent external force. A velocity dependent feedback control (VFC) has been recently implemented to reduce thermal noise of a cantilever in atomic force microscopy (AFM) \cite{LMC} and in dynamic force microscopy  \cite{THO}. A contribution to a rigorous thermodynamical foundation of macromolecules under (VFC) operating in nonequilibrium steady state is provided in \cite{KQ}. The entropy production rate is there decomposed into a {\em positive entropy production rate} (PEPR) and an {\em entropy pumping rate} (EPuR). The latter indicates how much entropy is pumped out or into the macromolecule by the control force. It may render the overall entropy production negative. This is at the basis of a macromolecular cooling mechanism \cite{LMC,THO,BBP}.

In this paper, we study entropy production in the presence of an
external force  in a more general situation. Our approach is new and complementary to  \cite{KQ} in that we study the free energy change rather than the total entropy change of the heat bath and of the Brownian particles as done in  \cite{KQ}. As ``distance"
between two probability densities associated to the unperturbed
and perturbed evolution we employ the information {\em relative
entropy} (in the quantum case, the  von Neumann relative entropy
for density operators). We show that it is possible to derive some
basic formulas on the entropy production rate that extend those of
\cite{KQ} whenever the evolution of the physical system is in some
suitable sense {\em Markovian}. We study  nonequilibrium
thermodynamical systems with finitely many degrees of freedom.
Corresponding results are also sketched  for $n$- level closed and
open quantum systems.  The perturbation  of the Hamiltonian is
interpreted as a {\em control function} which is designed by the
controller in order to obtain a desired behavior of the system
(reduction of thermal noise, transfer to another state, etc.).
Among potential applications, we mention  molecular kinetics
\cite{DPRR}, macromolecular cooling \cite{LMC,THO,BBP,KQ}, quantum
computation \cite{NC}. As it is well-known, relative entropy plays
a central role in many areas of modern science besides physics
such as  mathematical statistics, information theory, probability,
signal processing and quantum information processing, see e.g.
\cite{Ku,Ja,Cs,CT,Ge,GL,F2,Q,BA,CA,Pe,SW,Ve} and references
therein. Some of these results have been announced without  proofs
in our conference papers \cite{PT,PT2}.

The paper is outlined as follows. In the next section, we consider
finite dimensional, nonequilibrium thermodynamical systems. In
Section III, we derive a basic formula on relative entropy
evolution for probability densities satisfying a continuity type
equation. This result is then applied in the following section to
controlled thermodynamic systems. Section V is devoted to the
study of von Neumann entropy production for closed and open
finite-dimensional quantum systems. In the Appendix, we show that
the basic result may also be extended to {\em non-Markovian}
finite-energy diffusions.

\section{THERMODYNAMIC SYSTEMS}

Consider  an open thermodynamic system whose macroscopic evolution
is modelled by an n-dimensional Markov diffusion process
$\{x(t);t_0\le t\}$. The components of $x$ form a {\it complete
set} , i.e. all other variables have a much shorter relaxation
time \cite{Gr}.  Let $\bar{\rho}(x)$ be the Maxwell-Boltzmann
probability density corresponding to thermodynamical equilibrium
\begin{equation}\label{MB}
\bar{\rho}(x) = Z^{-1} \exp[ - \frac{H(x)}{kT}].
\end{equation}
Here $H$ is the (continuously differentiable) Hamiltonian
function, and the (forward) Ito differential of $x$ is
\begin{equation}\label{TH}dx(t)=\left[-\frac{1}{2kT}\Sigma\Sigma^T\nabla
H(x(t))+u(x(t),t)\right]dt+\Sigma dW,
\end{equation}
where $W$ is a standard $n$-dimensional Wiener process. The
probability density $\rho_t$ of $x(t)$ satisfies the {\it
Fokker-Planck equation}
\begin{equation}\label{FP1}
\frac{\partial{\rho}}{\partial{t}} +\nabla \cdot
\left[\left(-\frac{1}{2kT}\Sigma\Sigma^T\nabla
H+u\right)\rho\right] =
\frac{1}{2}\sum_{i,j=1}^n(\Sigma\Sigma^T)_{ij}\frac{\partial^2\rho}{\partial
x_i\partial x_j}.
\end{equation}
In the uncontrolled case $u=0$, under reasonable assumptions, see
e.g. \cite{Le}, as $t\rightarrow\infty$, the density $\rho_t$ of
$x(t)$ tends to $\bar{\rho}$ in {\em relative entropy} and,
consequently, in total  variation \cite{LC}.  For the ergodic
properties of this class of diffusions see e.g. \cite[Section
7.5]{St}.

\begin{example}
{\em In polymer dynamics \cite{DE}, the {\em macromolecule} is
described by a Hamiltonian
$$H(x,y)=\frac{1}{2}\langle y,My\rangle+\varphi(x),$$
where $M$ stands for the direct sum
$$M=M_1\oplus\cdots\oplus M_N, \quad M_k=m_k I_3,\quad k=1,\ldots,N,
$$
Here $x$ and $y$ are $3N$-dimensional vectors, with $x_i$, $y_i$
3-dimensional position and momentum of $i$th hard building block
of macromolecule. Moreover, $\varphi(x)$ is the {\em internal
potential} of macromolecule (in AFM experiment \cite{LMC},
$\varphi(x)=Kx^{2}/2$, where $K$ is the spring constant of cantilever).
Random collisions between solvent water molecules and building
blocks of macromolecule are modeled by the formal derivative of a
Wiener process, namely  {\em Gaussian white noise}. The six
dimensional stochastic process
$(q_{i,x},q_{i,y},q_{i,z},p_{i,x},p_{i,y},p_{i,z})$ associated the
$i$th block obeys the equation
\begin{eqnarray} \label{OU1} dq_{i\alpha}&=& \partial_{p_{i\alpha}} H(q,p)dt,\\ dp_{i\alpha}&=&\left[-\partial_{q_{i\alpha}} H(q,p)+f_{i\alpha}+u_{i\alpha}(q,p)\right]dt+\Gamma^{j\beta}_{i\alpha}dW_{j\beta}(t),\label{OU2}
\end{eqnarray}
where Einstein's convention has been used. Here $f$ is a
frictional force, $u$  a {\em position-velocity dependent
control}. In the AFM experiment, $f=-\gamma V$, $u=-\alpha V$,
$\gamma>0$, $\alpha>0$, with $V$ a velocity. The control here acts
{\em like} a frictional force on the macromolecule. Since the
frictional coefficient has been increased, one can introduce an
{\em effective} temperature $T_{{\rm eff}}$  which is lower than
the thermostat temperature $T$. As is well-known, different
uncontrolled ($u=0$) versions of this model \cite{HP1} play an
important role also in other applications such as nonlinear
circuits with noisy resistors \cite{TW}.}
\end{example}

To simplify the writing, we shall assume henceforth that in
(\ref{TH}) $\Sigma\Sigma^T=\sigma^2 I_n$. The results of this
paper, however,  extend in a straightforward way to the case
where the diffusion matrix $\Sigma\Sigma^T$ is any symmetric, non negative definite (possibly singular as in (\ref{OU1})-(\ref{OU2})) matrix. Let us first recall a few basic
concepts concerning the uncontrolled, nonequilibrium system
(\ref{TH}).Let us  introduce the {\em fluxes} $J(x,t)$ and {\em
forces} $\Phi(x,t)$ by
\begin{eqnarray}\nonumber J(x,t)&=&-\frac{1}{2}\sigma^2\nabla\rho_t(x)-\frac{1}{2kT}\sigma^2\nabla
H(x)\rho_t(x)\\\Phi(x,t)&=&-\nabla\mu(x,t),\nonumber
\end{eqnarray}
where $\mu=H+kT\log\rho_t$ is  the {\em electrochemical
potential}. Notice the following:
\begin{enumerate}
\item The Fokker-Plank equation (\ref{FP1}) may be rewritten (see
e.g. \cite{Gr}) as a continuity equation
\begin{equation}\label{COE}
\frac{\partial{\rho}}{\partial{t}} +\nabla \cdot J=0.
\end{equation}
\item Both fluxes and forces are zero in equilibrium. Moreover,
\begin{equation}\label{FL}J(x,t)=\frac{\sigma^2}{2kT}\Phi(x,t)\rho_t(x),
\end{equation}
which plays the role of {\em constitutive relations}. \item For
$\rho$ and $\sigma$ nonnegative measurable functions on $\R^n$, we
define the information {\em relative entropy} ({\em divergence,
Kullback-Leibler distance}) by
$$\D(\rho||\sigma)=\int_{\R^n}\log\frac{\rho}{\sigma}\rho \,dx.
$$
As it is well known \cite{Ku}, when $\rho$ and $\sigma$ are
integrable functions with
$$\int_{\R^n}\rho(x)dx=\int_{\R^n}\sigma(x)dx,
$$
we have $\D(\rho||\sigma)\ge 0$. Moreover, $\D(\rho||\sigma)=0$ if
and only if $\rho=\sigma$. Define the  {\em free energy}
functional
$$F(\rho_t)=kT\int_{\R^{n}}\log\frac{\rho_t}{\bar{\rho}}\rho_t
\,dx=kT\D(\rho_t||\bar{\rho}).
$$
The free energy decay may now be expressed as \cite{Gr}
\begin{equation}\label{FE}
\frac{d}{dt}F(\rho_t)=-\frac{\sigma^2kT}{2}
\int_{\R^n}|\nabla\log\frac{\rho_t}{\bar{\rho}}|^2\rho_t\,dx=-\int
J(x,t)\Phi(x,t)dx.
\end{equation}
\end{enumerate}
Suppose now that, like in the AFM experiment,  the thermodynamic
system is subject to a feedback control action so that the
macroscopic evolution is given by (\ref{TH}) with $u\neq 0$. The
density $\rho^u_t$ of the solution $x^u_t$ satisfies the
controlled Fokker-Planck equation (\ref{FP1}). We are interested
in the evolution of $\D(\rho^u_t||\rho^0_t)$, where $\{\rho^0_t,
t\ge t_0\}$ is an uncontrolled evolution ($u\equiv 0$) .  We first
need a simple but useful result.

\section{A relative entropy production formula}
Consider two families of nonnegative functions on $\R^n$ :
$\{\rho_t ;t_0\le t\le t_1\}$ and $\{\tilde{\rho}_t ;t_0\le t\le
t_1\}$. We are interested in how the relative entropy
$\D(\tilde{\rho_t}||\rho_t)$ evolves in time.
\newpage

\noindent {\bf Assumptions:}
\begin{itemize}
\item {\bf A1} There exist measurable functions $f(x,t)$ and
$\tilde{f}(x,t)$ such that $\{\rho_t ;t_0\le t\le t_1\}$ and
$\{\tilde{\rho}_t ;t_0\le t\le t_1\}$ are everywhere positive
$C^1$ solutions of
\begin{eqnarray}\label{C1}\frac{\partial \rho_t}{\partial t}+\nabla\cdot
(f\rho_t)=0,
\\\frac{\partial \tilde{\rho}_t}{\partial t}+\nabla\cdot
(\tilde{f}\tilde{\rho}_t)=0.\label{C2}
\end{eqnarray}
\item {\bf A2}

For every $t\in [t_0,t_1]$
\begin{eqnarray}\nonumber&&\lim_{|x|\rightarrow\infty}f(x,t)\tilde{\rho}_t(x)=0,\\\nonumber
&&\lim_{|x|\rightarrow\infty}\tilde{f}(x,t)\tilde{\rho}_t(x)=0,\\
&&\lim_{|x|\rightarrow\infty}\tilde{f}(x,t)\tilde{\rho}_t(x)\log
\frac{\tilde{\rho}_t}{\rho_t}(x)=0.\nonumber
\end{eqnarray}
\end{itemize}

\begin{theorem} \label{T1} Suppose $\D(\tilde{\rho_t}||\rho_t)<\infty ,
\forall t\ge 0$. Assume moreover A$1$ and A$2$ above. Then
$$
\frac{d}{dt}\D(\tilde{\rho_t}||\rho_t)=
\int_{\R^n}\left[\nabla\log\frac{\tilde{\rho}_t}{\rho_t}\cdot
(\tilde{f}-f)\right]{\tilde{\rho}_t}\,dx.
$$
\end{theorem}

\proof
\begin{eqnarray}\nonumber
\frac{d}{dt}\D(\tilde{\rho_t}||\rho_t)&=& \int_{\R^n}\frac{d}{dt}
\left[(\log
\tilde{\rho}_t-\log\rho_t)\tilde{\rho}_t\right]dx\\\nonumber
&=&\int_{\R^n}\left\{\left[\frac{1}{\tilde{\rho}_t} \frac{\partial
\tilde{\rho}_t}{\partial t}-\frac{1}{\rho_t} \frac{\partial
\rho_t}{\partial t}\right]\tilde{\rho}_t+
\log\frac{\tilde{\rho}_t}{\rho_t}\,\frac{\partial
\tilde{\rho}_t}{\partial t}\right\}dx\\\nonumber
&=&\int_{\R^n}\left[-\nabla\cdot(\tilde{f}\tilde{\rho}_t)+
\frac{\tilde{\rho}_t}{\rho_t}\, \nabla\cdot
(f\rho_t)-\log\frac{\tilde{\rho}_t}{\rho_t}\,\nabla\cdot
(\tilde{f}\tilde{\rho}_t)\right]dx\\&=&\int_{\R^n}\left[\nabla\log\frac{
\tilde{\rho}_t}{\rho_t}\cdot
\tilde{f}\tilde{\rho}_t-\nabla\frac{\tilde{\rho}_t}{\rho_t}\cdot
f\frac{\rho_t}{\tilde{\rho}_t}
\tilde{\rho}_t\right]dx=\int_{\R^n}\left[\nabla\log\frac{\tilde{\rho}_t}{\rho_t}\cdot
(\tilde{f}-f)\right]\tilde{\rho}_tdx,\nonumber
\end{eqnarray}
where we have used (\ref{C1})-(\ref{C2}) and integration by parts
(the boundary terms are zero because of Assumption $2$). Q.E.D.

\noindent For $\rho_t(x)\equiv 1$, we have that
$-\D(\tilde{\rho}_t||\rho_t)=S(\tilde{\rho}_t)$ the entropy.
Taking $f(x,t)\equiv 0$, we see that the first condition in A$2$
is verified. Theorem \ref{T1} then gives (exchanging $\rho_t$ with
$\tilde{\rho}_t$):
\begin{corollary} Suppose $\{\rho_t;t_0\le t\le  t_1\}$ is of
class $C^1$ and
$$\lim_{|x|\rightarrow\infty}v(x,t)\rho_t(x)=0,
\lim_{|x|\rightarrow\infty}v(x,t)\rho_t(x)\log \rho_t(x)=0.
$$
Suppose $S(\rho_t)<\infty, \forall t\in [t_0,t_1]$. Then
\begin{equation}\frac{d}{dt}S(\rho_t)=
-\int_{\R^n}\left[\nabla\log \rho_t\cdot f\right]\rho_t\,dx.
\end{equation}
\end{corollary}

\section{Entropy production for controlled evolution}

Consider now again the controlled  thermodynamic system of
Section II (\ref{TH}). Let $\rho^u_t$ denote the density of the
controlled process satisfying (\ref{FP1}). We are interested in
the evolution of $\D(\rho^u_t||\rho^0_t)$, where $\{\rho^0_t, t\ge
t_0\}$ is an uncontrolled evolution ($u\equiv 0$). First of all,
recall that the Fokker-Planck equations of the uncontrolled and
controlled system may be written as continuity equations as in
(\ref{COE}) Thus, we can apply Theorem \ref{T1} with
$$f=-\frac{\sigma^2}{2kT}\nabla
H(x)-\frac{\sigma^2}{2}\nabla\log\rho^0_t(x),\quad
\tilde{f}=-\frac{\sigma^2}{2kT}\nabla H(x)+
u(x,t)-\frac{\sigma^2}{2}\nabla\log\rho_t^u(x).$$ We get
\begin{equation}
\frac{d}{dt}\D(\rho^u_t||\rho^0_t)=
\int_{\R^n}\left(\nabla\log\frac{\rho^u_t}{\rho^0_t}\cdot
(u-\frac{\sigma^2}{2}\nabla\log\frac{\rho^u_t}{\rho^0_t}
)\right){\rho^u_t} dx. \label{CF}
\end{equation}
Suppose now  $\rho^0_t\equiv\bar{\rho}$, where $\bar{\rho}$ is the
Maxwell-Boltzmann distribution (\ref{MB}). We get
\begin{theorem} Under assumptions A$1$ and
A$2$,
\begin{equation}\label{CF1}
\frac{d}{dt}\D(\rho^u_t||\bar{\rho})=
-\frac{\sigma^2}{2}\int_{\R^n}\|\nabla\log\frac{\rho^u_t}{\bar{\rho}}\|^2{\rho^u_t}
dx+\int_{\R^n}\nabla\log\frac{\rho^u_t}{\bar{\rho}}\cdot
u{\rho^u_t} dx.
\end{equation}
\end{theorem}
\begin{remark} {\em Formula (\ref{CF1})   generalizes the decomposition of the entropy production exhibited in \cite{KQ} for the controlled Langevin equations.  In \cite{KQ}, the total entropy change of the heat bath and of the Brownian macromolecules is studied. The entropy production rate (EPR) is decomposed into the sum of two terms. The first, named PEPR ({\em positive entropy production rate}), is an always positive term  expressed as the product of the thermodynamic force and the corresponding flux as in (\ref{FE}). The second, named EPuR ({\em entropy pumping rate}), describes the amount of entropy pumped out of or into the macromolecule by the external agent.  We recognize that (\ref{CF1}) implies that $-\frac{d}{dt}\D(\rho^u_t||\bar{\rho})$ is also decomposed into an always positive term and into a term depending explicitly on  the control function.}
\end{remark}
One can try to employ (\ref{CF1}) to analize  macromolecular
cooling \cite{LMC,THO,BBP,KQ}. Another direction of application is
the following. Suppose we are interesting in modifying the rate at
which the solution $\rho_t$ of (\ref{FP1}) tends to the invariant
density (\ref{MB}). Let
$$\alpha(t)> -\frac{\sigma^2}{2},
$$
and consider in (\ref{TH}) the feedback control
\begin{equation}\label{FC}
u(x,t)=-\alpha(t) \nabla\log\frac{\rho^u_t}{\bar{\rho}}(x).
\end{equation}
Then, $\rho^u_t$ satisfies the Fokker-Planck equation
\begin{equation}\label{FP2}
\frac{\partial{\rho^u}}{\partial{t}} -\nabla \cdot
\left(\left(\frac{\sigma^2}{2}+\alpha(t)\right)\frac{1}{kT}\nabla
H\rho^u\right) =
\left(\frac{\sigma^2}{2}+\alpha(t)\right)\Delta\rho^u.
\end{equation}
A few observations are now in order.
\begin{enumerate}
\item Although the feedback control law is {\em nonlinear} in
$\rho^u_t$, equation  (\ref{FP2}) is {\em linear}; \item the
initial value problem for equation (\ref{FP2}) is well posed since
$\frac{\sigma^2}{2}+\alpha(t)>0 $; \item equation (\ref{FP2})
still has as invariant density the Maxwell-Boltzmann distribution
(\ref{MB}); \item it is conceivable to solve (\ref{FP2}) off-line,
and consequently compute the feedback law (\ref{FC}) beforehand.
\item the flow of one dimensional probability densities
$\{\rho^u_t; t\ge 0\}$ of $x^u(t)$ satisfying (\ref{TH}) with the
control given by (\ref{FC}) is the same as for the uncontrolled
stochastic process $\xi$ with differential
\begin{equation}\label{X} d\xi=-\left(\frac{\sigma^2}{2}+\alpha(t)\right)\frac{1}{kT}\nabla H(\xi) dt+\sqrt{\sigma^2+2\alpha(t)}dW,
\end{equation}
provided $\xi(0)$ is distributed according to $\rho^u_0$. \item
the friction and diffusion coefficients in (\ref{X}), although
time-varying, still satisfy Einstein fluctuation-dissipation
relation, see e.g \cite{N1}.
\end{enumerate}
We  now employ (\ref{CF1}) to compute the relative entropy
derivative.  We get
\begin{equation}
\frac{d}{dt}\D(\rho^u_t||\bar{\rho})=
-\left(\frac{\sigma^2}{2}+\alpha(t)\right)
\int_{\R^n}|\nabla\log\frac{\rho^u_t}{\bar{\rho}}|^2 \rho^u_t\,dx.
\end{equation}
Hence, the controlled diffusion still tends to the
Maxwell-Boltzmann distribution but at a different, ``modulated"
rate. In the linear Gauss-Markov  case (i.e. when $H$ is
quadratic), the results assume a very concrete form. In
particular, equation (\ref{FP2}) may be replaced by a linear
matricial equation, see \cite{PT} for details.

The results of this section may be readily extended to
non-Markovian, finite-energy diffusions employing the
Nelson-F\"{o}llmer kinematics \cite{N2,F}, see the Appendix.
Notice that this family  family plays a central role in several
branches of mathematical physics, see e.g. \cite{F2,N3}. The
results also extend without too much difficulty to a large class
of diffusions with constant but singular diffusion coefficient
such as in the case of the Orstein-Uhlenbeck model of physical
Brownian motion \cite{N1} or, more generally, in the case of model
(\ref{OU1})-(\ref{OU2}). They may also be established for a large
class of Markovian diffusion processes with local diffusion
coefficient given the results in \cite{N0,Na,Mo,N1,HP}.

\section{$n$-level Quantum Systems}
It is apparent that Theorem \ref{T1} can be applied to statistical
mixtures in classical mechanics \cite[Section IV]{PT2}. Indeed,
Liouville's equation, expressing conservation of density in phase
space, is just a continuity equation for the Hamiltonian
evolution. One then gets the idea that it might be possible to
establish a similar result  in the quantum case, replacing the
Liouville equation with the Landau-von Neumann equation for the
density operator. First of all, we need to recall the basic
formalism of statistical quantum mechanics.
\subsection{Closed quantum systems}
As in standard quantum mechanics \cite{sakurai}, to every physical
system $S$ is associated a complex Hilbert space $\mathcal{H}_S$.
In the standard formulation, the state of the system is described
by a unit vector
 $\psi\in \mathcal{H}_S$. For the sake of simplicity, here we will
consider only finite dimensional Hilbert spaces, but results hold
in the general case.

We consider situations in which uncertainty on the system state
affects our model. The quantum analogue of a classical probability
density is a density operators $\rho$ in $\mathcal{H}_S$: A
density operator is a a positive semi-definite, unit trace
operator  on $\mathcal{H}_S$. They form a convex set
$\mathcal{D}(\mathcal{H}_S)$ and the extremals of
$\mathcal{D}(\mathcal{H}_S)$ are the one dimensional orthogonal
projections. These are called \emph{pure states}, and are
equivalent to unit vectors in $\mathcal{H}_S$ up to an overall
phase factor, by setting $\rho=\langle\psi,\cdot\rangle\psi$.
Physical observables are represented by Hermitian operators on
$\mathcal{H}_S$.

Let $A$ be an observable: The expected value of $A$ for a system
described by a density operator $\rho$ is defined as:
\begin{equation}
<a>_\rho:=\tr(\rho A). \end{equation} Hence, the variance for an
observable $A$ given $\rho$ is naturally defined as
\begin{equation} \textrm{Var}(A)_\rho:=<(A-<a>_\rho)^2>_\rho. \end{equation}
It is easy to see that if $\rho_p$ is a pure state, then exists an
observable $A$ such that the variance
$\textrm{Var}(A)_{\rho_p}=0,$ clarifying the definition and the
analogy with the classical case. The time evolution for the
density operator of an isolated quantum system is determined by
the Hamiltonian, i.e. the energy observable. The dynamical
equation is the  \emph{Landau- von Neumann equation}:

\begin{equation}\label{me}
   i\hbar\frac{d}{dt}\rho_t=[H,\rho_t],
   \end{equation}
where $[\cdot,\cdot]$  denotes the commutator
$$[A,B]:=AB-BA,$$
and $\hbar$ is Planck's constant divided by $2\pi$.

Quantum analogues of entropic functionals have been considered
since the very beginning of the mathematical foundation of quantum
mechanics \cite{vonneumann}. Recently renewed interest came from
Quantum Information applications \cite{NC}. We are interested here
in the quantum relative entropy, that is defined as:
\begin{equation}
\D(\rho||\tilde{\rho}):=\tr(\rho(\log \rho -\log \tilde{\rho})).
\end{equation}
We define
$0\log 0=0$. As in the classical case, quantum relative entropy
has the property of a pseudo-distance (see e.g. \cite{NC,SW}).
We now consider the effect of a perturbation $\Delta H$
on the evolution of quantum system originally driven by a free Hamiltonian
$H$ (we denote by $\tilde{H}=H+\Delta H$ the perturbed Hamiltonian) .
\begin{proposition}  Let $\rho_t$ and $\tilde{\rho}_t$ be the solution of (\ref{me})
corresponding to the unperturbed and the perturbed Hamiltonians,
respectively. The relative entropy production for the perturbed
evolution is given by:
\begin{equation}\label{QREC} \frac{d}{dt}
\D(\rho||\tilde{\rho})=\frac{i}{\hbar}<[\Delta
H,\log\tilde{\rho}]>_\rho.
\end{equation}
\end{proposition}

\proof Observing that $[\rho,\log\rho]=0$ and, consequently,
$$\frac{d}{dt}
\tr({\rho}\log{\rho})=0,$$ 
(i.e. the \emph{von Neumann entropy} is
time invariant under Hamiltonian evolution), and using the cyclic
property of trace, we have:
\begin{eqnarray} \frac{d}{dt}
\D(\rho||\tilde{\rho})&=&\frac{d}{dt} \tr({\rho}\log{\rho})+\frac{i}{\hbar}\tr\left([{H},\rho]\log\tilde{\rho}+\rho[\tilde{H},\log\tilde{\rho}]\right)\nonumber\\
&=&\frac{i}{\hbar}\tr\left( [H,\rho\log\tilde{\rho}]+\rho[\Delta
H,\log\tilde{\rho}]\right)\nonumber\\
&=&\frac{i}{\hbar}\tr\left(\rho[\Delta H,\log\tilde{\rho}]\right).
\end{eqnarray} Q.E.D.\\

\noindent
We remark that the initial conditions for the perturbed and the
unperturbed evolution can be different, and we can easily exchange
the role of perturbed and unperturbed evolution adding a minus
sign on the right hand side. The analogy with the corresponding
relative entropy evolution formula in classical mechanics
\cite{PT2} is apparent. As in the classical case, the perturbation
can be interpreted as an additive control Hamiltonian.

\subsection{Open Quantum Systems}

When we consider a quantum system interacting with the environment
in some uncontrollable way, namely an \emph{open quantum system}
\cite{NC,alicki}, the situation changes significantly. The
complete dynamical description of the situation should be done
considering the tensor product space of both the system and the
environment space. Usually, the environment has too many degrees
of freedom to be modelled. Moreover, only partial information
about environment initial state interactions may be available. In
these cases, we can still obtain a dynamical equation for the
system state by averaging over the environment degrees of freedom
\cite{alicki}. If the system evolution is assumed to be Markovian,
strongly continuous in time and completely positive \cite{NC}, a
general form for the generator of the system density operator
dynamics is the following \cite{lindblad}:

\begin{equation}\label{lme}
\frac{d}{dt}\rho_t=-\frac{i}{\hbar}[H,\rho_t]+\mathcal{L}[\rho_t],
\end{equation}

\noindent where $H$ is the effective Hamiltonian, in general
different from the free drift Hamiltonian, and the generator for
the dissipative evolution $\mathcal{L}$ has the form:

\begin{equation}
\mathcal{L}[\rho]=\frac{1}{2}\sum_k\left([L_k\rho,L_k^\dag]+[L_k,\rho
L_k^\dag]\right).
\end{equation}
\noindent The operators $L_k$ can be derived under different
assumptions on the couplings with the environment or on a
phenomenological basis (see e.g. \cite{alicki} and reference
therein). This equation can be seen as a quantum analogue of a
Fokker-Planck equation, since it describes the time evolution of
the density operator in the absence of conditioning measurements.
Assume that (\ref{lme}) admits a stationary state commuting with
the effective Hamiltonian, and denote it with $\bar\rho$. Noting
that:$$\tr{\left(\rho\frac{d}{dt}\log\rho\right)}=\tr{\left(\frac{d}{dt}\rho\right)}=0,$$
since the generator (\ref{lme}) has zero trace, we obtain for the
relative entropy production (see also \cite{spohn}):

\begin{eqnarray}
\frac{d}{dt}\D(\rho||\bar{\rho})&=&\tr(\mathcal{L}[\rho](\log\rho
-\log\bar{\rho}))\le 0.
\end{eqnarray}
\noindent The fact that $\D(\rho||\bar{\rho})$ is non-increasing for the dynamical semigroup
generated by (\ref{lme}) was established by Lindblad, see e.g.
\cite{alicki}. To extend this result to the case of a perturbed Hamiltonian, we consider now $\bar\rho$ as a {\em fixed target
state}, since the introduction of perturbations could in general
change the stationary states.
In this setting, we get:
\begin{eqnarray}\label{QRECD}
\frac{d}{dt}\D(\rho||\bar{\rho})&=&\tr(\frac{i}{\hbar}[\tilde{H},\rho]\log\bar\rho+\mathcal{L}[\rho](\log\rho-\log\bar\rho))\nonumber\\
&=&-\frac{i}{\hbar}\left\langle[\Delta
H,\log{\bar\rho}]\right\rangle_{\rho}+\tr(\mathcal{L}[\rho](\log\rho
-\log\bar{\rho})),
\end{eqnarray}
where now $\rho$ is undergoing a \emph{perturbed evolution}
$\tilde{H}=H+\Delta H$ and $[\bar\rho,H]=0$ as before.
In the quantum case, however, the effectiveness of a  control Hamiltonian
is severely limited. For instance, in the closed system case, the density operator
eigenvalues cannot be modified by a control Hamiltonian,
precluding convergence in relative entropy if the target state has
a different spectrum from the initial condition. A detailed
analysis of the dissipative case from a control theoretic
viewpoint can be found e.g. in  \cite{altafini,schirmer}. Whether
these formulas could be of help in designing or analyzing control
strategies will be a matter of further work (see also comments on
this issue in \cite[Section VII]{PT}).

Further analogies with the classical thermodynamics setting can be
unravelled if we restrict our attention to equation (\ref{lme})
when it is derived from e.g. a \emph{weak coupling} limit
\cite{alicki}. This is essentially a constructive derivation of
equations of the form (\ref{lme}) from the joint (tensor)
description of the system and the environment, that is consistent
with classical thermodynamics. In fact, the Gibbs state:
$$\rho_G=Z^{-1}e^{-\beta H},$$ where $Z$ is the {\em partition
function} and $H$ the system Hamiltonian, is a stationary state
for the resulting equation. Since, as we already recalled,  relative entropy with respect to
the stationary state is non-increasing, for this class of dissipative Markovian evolutions
we have a full correspondence with the classical mechanical case
\cite{PT2}.

\section{Conclusion and outlook}
We have derived explicit dependence of  the {\em relative entropy
production rate} on the control action for various uncertain
physical systems exhibiting a  {\em Markovian evolution}. Further
work is needed to find other significant applications of the
results as well as possible extension to other, more complex,
systems with Markovian evolution such as interacting particle
systems.

\appendix*
\section{Finite-energy diffusions}

Let $\Omega:={\cal C}([t_0,t_1],\R^n)$ denote the family of
$n$-dimensional continuous functions, and let $P$ and $\tilde{P}$
be two probability distributions on $\Omega$. The {\it relative
entropy } $H(\tilde{P},P)$ of $\tilde{P}$ with respect to $P$ is
defined by
$$H(\tilde{P},P)=\left\{\begin{array}{ll}
E_{\tilde{P}}[\log\frac{d\tilde{P}}{dP}] &\mbox{if
$\tilde{P}<<P$}\\ +\infty & \mbox{otherwise}\end{array}\right.$$
Let $W_x$ denote Wiener measure on $\Omega$ starting at
$x\in\R^n$, and let
$$W:=\int W_x\,dx
$$
be stationary Wiener measure. Let $\sigma>0$, and denote by $\D$
be the family of distributions $P$ on $\Omega$ such that
$H(P,\sigma W)<\infty$. Let ${\cal F}_t$ and ${\cal G}_t$ denote
the $\sigma$-algebras of events observable up to time $t$ and from
time $t$ on, respectively. It then follows from the Girsanov's
theory \cite{F,KS} that $P\in\D$ possesses both a forward drift
$\beta^P$ and a backward drift $\gamma^P$, namely under $P$, the
increments of the canonical coordinate process
$x(t,\omega)=\omega(t)$ admit the representations
\begin{eqnarray} x(t)-x(s)=\int_s^t\beta^P(\tau)d\tau+
\sigma[w_+(t)-w_+(s)],\quad t_0\le s<t\le t_1,\\
x(t)-x(s)=\int_s^t\gamma^P(\tau)d\tau+ \sigma[w_-(t)-w_-(s)],\quad
t_0\le s<t\le t_1.
\end{eqnarray}
$\beta^P(t)$ is at each time $t$ ${\cal F}_t$-measurable and
$w_+(\cdot)$ is a standard, n-dimensional Wiener process.
Symmetrically,  $\gamma^P(t)$ is ${\cal G}_t$-measurable and $w_-$
is another standard Wiener process. Moreover, $\beta^P$ and
$\gamma^P$ satisfy the finite-energy condition
\begin{equation}\label{K2}
E\left\{\int_{t_0}^{t_1}\beta^P(t)\cdot\beta^P(t)dt\right\}<\infty,
\quad
E\left\{\int_{t_0}^{t_1}\gamma^P(t)\cdot\gamma^P(t)dt\right\}<\infty.
\end{equation}
It was shown in \cite{F} that the one-time probability density
$p_t(\cdot)$ of $x(t)$ (which exists for every $t$) is absolutely
continuous on $\R^n$ and the following relation holds a.s.
$\forall t>0$ \begin{equation}\label{K4'}
E\{\beta^P(t)-\gamma^P(t)|x(t)\} = \sigma^2\nabla\log p_t(x(t)).
\end{equation} Let us introduce the {\em current drift} and the
{\em current drift field} of $P$
\begin{equation} \label{CU}v^P(t) = \frac{\beta^P (t) + \gamma^P
(t)}{2},\quad v^P(x,t)=E\{v^P(t)|x(t)=x\}.
\end{equation}
Then, the one-time density $p_t$  satisfies weakly \cite{N2} a
continuity type equation
\begin{equation}
\frac{\partial p_t}{\partial t}+\nabla\cdot (v^Pp_t)=0.
\end{equation}
Hence, Theorem \ref{T1} holds true for finite energy diffusions
provided we define the $v$ fields according to (\ref{CU}).

\end{document}